\documentclass[10pt,letterpaper]{article}
\usepackage{geometry}
\usepackage{graphicx}

\usepackage{amsmath,amssymb}
 \usepackage{changepage}
\usepackage[utf8x]{inputenc}
\usepackage{textcomp,marvosym}
\usepackage{cite}
\usepackage{nameref,hyperref}
\usepackage{microtype}
\DisableLigatures[f]{encoding = *, family = * }

\usepackage[table]{xcolor}

\usepackage{array}

\newcolumntype{+}{!{\vrule width 2pt}}

\newlength\savedwidth

\usepackage[aboveskip=1pt,labelfont=bf,labelsep=period,justification=raggedright,singlelinecheck=off]{caption}

\makeatletter
\renewcommand{\@biblabel}[1]{\quad#1.}
\makeatother

\date{}

\newcommand{\gen}{t}                    
\newcommand{\ngen}{n_{\text{gen}}}      
\newcommand{\mgen}{m_{\text{gen}}}      
\newcommand{\nrem}{n_{\text{rem}}}      
\newcommand{\Krem}{K_{2}^{\text{rem}}}  
\newcommand{\df}{D_{\text{f}}}          
\newcommand{\pc}{p_{\text{c}}}          
\newcommand{\oc}{\omega_{\text{c}}}     
\newcommand{\crem}{C_{\text{rem}}}      
\newcommand{\DR}{\Delta_{\text{R}}}     
\newcommand{\DRa}{\Delta_{\text{R1}}}   
\newcommand{\DRb}{\Delta_{\text{R2}}}   
\newcommand{\Dgen}{\Delta_{\text{gen}}} 

\begin{document}

\vspace*{0.2in}

\begin{flushleft}

{\Large A General Model of Hierarchical Fractal Scale-free Networks }
\newline
\\
Kousuke~Yakubo\textsuperscript{1*}, Yuka~Fujiki\textsuperscript{2},
\\
\bigskip
\textbf{1} Department of Applied Physics, Hokkaido University, Sapporo 060-8628, Japan
\\
\textbf{2} Advanced Institute for Materials Research, Tohoku University, Sendai 980-8577, Japan
\bigskip

* Corresponding author
E-mail: yakubo@eng.hokudai.ac.jp (KY)

Both contributed equally to this work.

\end{flushleft}
\section*{Abstract}
We propose a general model of unweighted and undirected networks having the scale-free property and
fractal nature. Unlike the existing models of fractal scale-free networks (FSFNs), the present
model can systematically and widely change the network structure. In this model, an FSFN is
iteratively formed by replacing each edge in the previous generation network with a small graph
called a generator. The choice of generators enables us to control the scale-free property,
fractality, and other structural properties of hierarchical FSFNs. We calculate theoretically
various characteristic quantities of networks, such as the exponent of the power-law degree
distribution, fractal dimension, average clustering coefficient, global clustering coefficient, and
joint probability describing the nearest-neighbor degree correlation. As an example of analyses of
phenomena occurring on FSFNs, we also present the critical point and critical exponents of the
bond-percolation transition on infinite FSFNs, which is related to the robustness of networks
against edge removal. By comparing the percolation critical points of FSFNs whose structural
properties are the same as each other except for the clustering nature, we clarify the effect of
the clustering on the robustness of FSFNs. As demonstrated by this example, the present model makes
it possible to elucidate how a specific structural property influences a phenomenon occurring on
FSFNs by varying systematically the structures of FSFNs. Finally, we extend our model for
deterministic FSFNs to a model of non-deterministic ones by introducing asymmetric generators and
reexamine all characteristic quantities and the percolation problem for such non-deterministic
FSFNs.

\section*{Introduction}
Many of the complex systems around us and in various research fields of science and technology can
be described by networks \cite{Albert02,Barthelemy11,Cimini19,Gosak18,Ding19}. Since nodes and
edges, the constituents of networks, represent a wide variety of objects and interactions,
respectively, Euclidean distances are not always defined for networks. The absence of the Euclidean
distance eliminates the limitation of the number of edges connecting to a node, namely the degree
$k$ of a node, and thus allows a large fluctuation of $k$. In fact, degree distributions $P(k)$ of
many real-world networks obey power-law functions for large $k$, i.e. $P(k)\propto k^{-\gamma}$
with an exponent $\gamma$ \cite{Barabasi99}. We define the shortest path distance to be the minimum
number of edges connecting two nodes even for a network where the Euclidean distance is not
defined. We can quantify the linear distance over a network by the average shortest-path distance
$\langle l \rangle$ or the network diameter $L$ defined as the largest shortest-path distance. If
the diameter $L$ of a network $\mathcal{G}$ scales with the number of nodes $N$ as $L\propto \log N$,
$\mathcal{G}$ is referred to as a small-world network \cite{Watts98}. In contrast, if the
relation $L\propto N^{1/\df}$ holds, $\mathcal{G}$ is called to be a \textit{fractal network} with
the fractal dimension $\df$ \cite{Song05}. Due to the small-world nature of the majority of
real-world networks, a lot of structural models of small-world and scale-free networks have been
proposed \cite{Barabasi99,Dorogovtsev00,Li03,Krapivsky05,Vazquez01,Boguna03,Caldarelli02,Barrat04,Newman10},
and various phenomena or dynamics on them have been extensively studied
\cite{Boccaletti06,Dorogovtsev08,Li21,Arenas08,Kiss18}. Actual scale-free networks, however, often
possess fractal structures in shorter length scales than their network diameters or the average
shortest-path distances. World Wide Web, protein interaction networks, and actor networks are known
to be examples of real-world fractal scale-free networks \cite{Song05,Kitsak07,Kawasaki10}.
Nevertheless, there is less research on fractal scale-free networks (FSFNs) than on small-world
scale-free networks. In particular, the lack of a structural model of FSFNs that can freely control
the exponent $\gamma$, fractal dimension $\df$, and other structural features delays the study of
phenomena or dynamics on FSFNs.

There are two representative models of FSFNs, the $(u,v)$-flower model \cite{Rozenfeld07a} and the
Song-Havlin-Makse (SHM) model \cite{Song06}. The $(u,v)$-flower model constructs hierarchically a
highly cycle-rich FSFN and can vary the network structure to some extent by adjusting the
parameters $u$ and $v$ ($u\ge v\ge 2$). The clustering coefficient is, however, always zero
independently of $u$ and $v$. Meanwhile, the SHM model forms a tree-like FSFN and can change the
fractal dimension $\df$ (and then $\gamma$ followed by the change in $\df$) by controlling the
parameter $z$ characterizing the growth rate of nodes. Since the network has a tree structure, the
clustering coefficient is also zero for any value of the parameter $z$ in the SHM model. In
addition to these two structural models, several deterministic models for FSFNs have been proposed
so far \cite{Zhang09,Pan10,Hu85,Zhang11b,Gao21,Zhang11,Kuang15}. These are, however, classified
into derivatives of the $(u,v)$-flower model or the SHM model, or synthetic models in which the
network structure cannot be freely controlled.

In this paper, we propose a general model of hierarchical FSFNs which can change widely and freely
the structural features of obtained networks. This model constructs an FSFN hierarchically by
replacing each edge in the previous generation network iteratively with a small graph called a
generator. By choosing the generator appropriately, it becomes possible to control the scale-free
nature, fractality, clustering property, and the nearest-neighbor degree correlation. We actually
calculate analytically the exponent $\gamma$ describing the scale-free property, fractal dimension
$\df$, average clustering coefficient $C$, global clustering coefficient $C^{\triangle}$, and joint
probability $P(k,k')$ defining the nearest-neighbor degree correlation for networks formed by the
present model. Furthermore, the bond-percolation problem on constructed FSFNs is investigated as an
example of analyses of phenomena occurring on them. Specifically, we analytically derive the
percolation critical point and critical exponents for infinitely large FSFNs, by using structural
information of the generator. In addition, a model of non-deterministic FSFNs is provided by
introducing asymmetric generators, and various statistical properties of resulting networks are
examined theoretically.

\section*{Model}
\begin{figure}[ttt]
\begin{center}
\includegraphics[width=13cm]{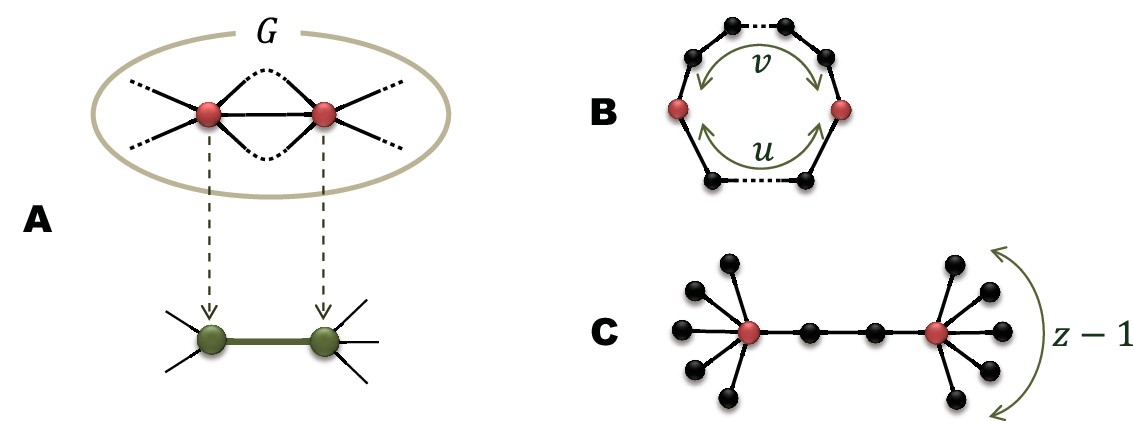}
\caption{{\bf Network formation by symmetric generators.} (A) Every edge in $\mathcal{G}_{\gen-1}$
is replaced with $G$ so that the terminal nodes of the edge (green nodes) coincide with the root
nodes of $G$ (red nodes). (B) Generator $G$ for the $(u,v)$-flower. (C) Generator $G$ for the SHM
model. Red nodes in (B) and (C) represent the root nodes of $G$.}
\label{fig1}
\end{center}
\end{figure}

We first prepare a small connected graph $G$ (called \textit{generator} hereafter) in which two
particular nodes are specified as \textit{root nodes}. As shown in Fig~\ref{fig1}A, a network in
the $\gen$-th generation, $\mathcal{G}_{\gen}$, is constructed by replacing every edge in
$\mathcal{G}_{\gen-1}$ with the generator $G$ so that the terminal nodes of the edge coincide with
the root nodes of $G$. This procedure is an inversion operation of the renormalization
transformation that replaces small subgraphs with superedges \cite{Rozenfeld07a,Kim07}. Although the
initial network $\mathcal{G}_{0}$ can be arbitrarily chosen, we fix, in this work, $\mathcal{G}_{0}$
to be a single edge connecting two nodes for simplicity. In order to obtain a deterministic fractal
scale-free network (FSFN), the generator $G$ must satisfy the following three conditions:
\begin{description}
\item[(1)] The degree of the root node is no less than $2$.
\item[(2)] The shortest-path distance between the two root nodes is $2$ or longer.
\item[(3)] The two root nodes are symmetric to each other in $G$.
\end{description}
The first and second conditions guarantee the scale-free property and the fractal nature of
the constructed network, respectively. If the first condition is violated and the degree of the
root node is one, our model produces a non-scale free network with an exponentially damped degree
distribution. Such networks can also be formed by existing models \cite{Nowotny88,Rosenberg20}.
In the case of the violation of the second condition, namely, when the two root nodes
are directly connected by an edge, our model gives hierarchical scale-free networks with the
small-world property such as a network modeled by \cite{Dorogovtsev02}. It should be emphasized
that even if the conditions (1) and (2) are violated, all the analytical arguments below still
hold, except for those in the sections ``Scale-free property", ``Fractal property", and ``Degree
correlation". The third condition is for the model to be deterministic. The meaning of ``symmetric"
in the third condition is the following. If a network constructed by removing one root node and
its edges from $G$ has the same topology as a network formed by removing another root node and
its edges from $G$, these root nodes are called to be symmetric.
%

The most distinct advantage of this model is that a generator $G$ can be chosen arbitrarily (within
the three conditions) and thus a wide variety of hierarchical FSFNs are produced by this model,
whereas previous models for FSFNs based on inverse renormalization procedure either fix $G$ or
limit the structural freedom of $G$ to a narrow range. Therefore, the present model is able to
construct FSFNs proposed by previous models \cite{Rozenfeld07a,Song06,Zhang09,Pan10,Hu85,Zhang11b,Gao21}.
For example, the $(u,v)$-flower \cite{Rozenfeld07a} is reproduced by choosing a generator $G$ of a
cycle with $u+v$ nodes as shown in Fig~\ref{fig1}B. If we start with a star graph with $4$ leaves
as the initial graph $\mathcal{G}_{0}$ instead of an edge and employ a generator consisting of two
connected star graphs with $z$ leaves as shown in Fig~\ref{fig1}C, we obtain the Song-Havlin-Makse
(SHM) model \cite{Song06}. Furthermore, by using a generator with the different number of edges
connecting the root nodes in Fig~\ref{fig1}C, we can generate the fractal scale-free tree proposed
by \cite{Zhang11b,Gao21}.

\section*{Results}
\subsection*{Properties of fractal scale-free networks}
Any characteristics concerning a constructed network $\mathcal{G}_{\gen}$ in the $\gen$-th
generation are completely determined by the nature of the generator $G$. Many of quantities
defining $\mathcal{G}_{\gen}$ can be analytically calculated because of the simplicity of the
model. We present here several indices of $\mathcal{G}_{\gen}$ by using quantities describing the
generator $G$.

\subsubsection*{Numbers of edges and nodes}

The most fundamental quantities are the numbers of nodes and edges in $\mathcal{G}_{\gen}$. The
number of edges $M_{\gen}$ in the $\gen$-th generation network $\mathcal{G}_{\gen}$ is $\mgen$
times larger than the number of edges in $\mathcal{G}_{\gen-1}$, namely,
\begin{equation}
M_{\gen}=\mgen M_{\gen-1} ,
\label{eq_3_1}
\end{equation}
where $\mgen$ is the number of edges in the generator $G$. This relation with $M_{0}=1$ immediately
leads to
\begin{equation}
M_{\gen}=\mgen^{\gen} .
\label{eq_3_2}
\end{equation}
The number of nodes $N_{\gen}$ in $\mathcal{G}_{\gen}$ is the sum of the number of nodes
$N_{\gen-1}$ in the previous generation network $\mathcal{G}_{\gen-1}$ and the number of newly
added nodes in the replacement of edges in $\mathcal{G}_{\gen-1}$. It is convenient for counting
newly added nodes and for later discussion to define \textit{remaining nodes} which are the nodes
in the generator $G$ other than the root nodes. For each edge in $\mathcal{G}_{\gen-1}$, the edge
replacement with $G$ introduces $\nrem(=\ngen-2)$ nodes, where $\ngen$ and $\nrem$ are the numbers
of the entire nodes and remaining nodes in $G$, respectively. The number of newly added nodes in
$\mathcal{G}_{\gen}$ is thus given by $M_{\gen-1}\nrem$ and $N_{\gen}$ is expressed by
$N_{\gen}=N_{\gen-1}+M_{\gen-1}\nrem$. Solving this recurrence relation with $N_{0}=2$ with the aid
of Eq~(\ref{eq_3_2}), the number of nodes in $\mathcal{G}_{\gen}$ is given by
\begin{equation}
N_{\gen}=2+\frac{\nrem(\mgen^{\gen}-1)}{\mgen-1} .
\label{eq_3_3}
\end{equation}
For $\gen\gg 1$, this equation leads to the approximate relation
\begin{equation}
N_{\gen}\approx \mgen N_{\gen-1} .
\label{eq_3_4}
\end{equation}
The quantities $M_{\gen}$ and $N_{\gen}$ are not influenced by the structure of the generator $G$
but are determined only by the numbers of nodes and edges in $G$. Many other indices characterizing
$\mathcal{G}_{\gen}$, however, depend on the topology of $G$ as shown below.

Let us consider the number of nodes $N_{\gen}(k)$ of degree $k$ in $\mathcal{G}_{\gen}$. A node of
degree $k$ in $\mathcal{G}_{\gen-1}$ has the degree $\kappa k$ in $\mathcal{G}_{\gen}$, where
$\kappa$ is the degree of the root node in $G$. Therefore, the number of nodes of degree $k$ in
$\mathcal{G}_{\gen}$ is the sum of the number of nodes of degree $k/\kappa$ in
$\mathcal{G}_{\gen-1}$ only if $k/\kappa$ is an integer and the number of degree $k$ nodes which
are newly added in the edge replacement operation. Thus, for $\gen\ge 1$, we have the relation,
\begin{equation}
N_{\gen}(k)=N_{\gen-1}(k/\kappa)+M_{\gen-1}\sum_{n\in G}^{\nrem}\delta_{k,k_{n}^{\text{rem}}}  ,
\label{eq_3_5}
\end{equation}
where $N_{\gen}(k/\kappa)=0$ if $k/\kappa$ is non-integer, the summation is taken over the $\nrem$
remaining nodes in $G$, $k_{n}^{\text{rem}}$ is the degree of the $n$-th remaining node, and
$\delta_{k,k'}$ is the Kronecker delta. Considering that $N_{0}(k)=2\delta_{k,1}$ for
$\mathcal{G}_{0}$ of a single edge, Eq~(\ref{eq_3_5}) gives
\begin{equation}
N_{\gen}(k)=2\delta_{k,\kappa^{\gen}}+
\sum_{\gen'=1}^{\gen}\sum_{n\in G}^{\nrem}\mgen^{\gen'-1}\delta_{k,\kappa^{\gen-\gen'}k_{n}^{\text{rem}}} .
\label{eq_3_6}
\end{equation}
We can calculate $N_{\gen}(k)$ if $\kappa$, $\mgen$, $\nrem$, and $k_{n}^{\text{rem}}$
characterizing the generator $G$ are given.

Using the above expression of $N_{\gen}(k)$, the average degree $\langle k\rangle_{\gen}$ and the
average squared degree $\langle k^{2}\rangle_{\gen}$ of $\mathcal{G}_{\gen}$ are evaluated. Since
it is obvious that the average degree must be given by $\langle
k\rangle_{\gen}=2M_{\gen}/N_{\gen}$, Eqs~(\ref{eq_3_2}) and (\ref{eq_3_3}) lead to
\begin{equation}
\langle k\rangle_{\gen}=\frac{2\mgen^{\gen}(\mgen-1)}{2(\mgen-1)+\nrem(\mgen^{\gen}-1)}  .
\label{eq_3_7}
\end{equation}
This can be confirmed by calculating $\sum_{k}kN_{\gen}(k)/N_{\gen}$ with the aid of
Eq~(\ref{eq_3_6}). In the limit of infinite size $N_{\gen}$ ($\gen \to \infty$), the average degree
is then given by
\begin{equation}
\langle k\rangle_{\infty}=\frac{2(\mgen-1)}{\nrem}  .
\label{eq_3_8}
\end{equation}
The average squared degree $\langle k^{2}\rangle_{\gen}=\sum_{k}k^{2}N_{\gen}(k)/N_{\gen}$ is
obtained by using Eq~(\ref{eq_3_6}) as
\begin{equation}
\langle k^{2}\rangle_{\gen}=\frac{2 \kappa^{2\gen}}{N_{\gen}}+\frac{\kappa^{2\gen}\Krem}{N_{\gen}\mgen}
\sum_{\gen'=1}^{\gen}\left(\frac{\mgen}{\kappa^{2}}\right)^{\gen'}  ,
\label{eq_3_9}
\end{equation}
where
\begin{equation}
\Krem=\sum_{n\in G}^{\nrem}(k_{n}^{\text{rem}})^{2}  .
\label{eq_3_10}
\end{equation}
Taking the summation over $\gen'$ in Eq~(\ref{eq_3_9}) and using Eq~(\ref{eq_3_3}) for $N_{\gen}$,
$\langle k^{2}\rangle_{\gen}$ is expressed as
\begin{equation}
\langle k^{2}\rangle_{\gen}=
\begin{cases}
\displaystyle \frac{(\mgen-1)\left[ 2\kappa^{2\gen}(\mgen-\kappa^{2})+\Krem(\mgen^{\gen}-\kappa^{2\gen})\right] }
{(\mgen-\kappa^{2})\left[2(\mgen-1)+\nrem(\mgen^{\gen}-1)\right]} & \text{for } \mgen\ne \kappa^{2} \\[14pt]
\displaystyle \frac{\kappa^{2(\gen-1)}(\kappa^{2}-1)(2\kappa^{2}+\gen\Krem)}
{2(\kappa^{2}-1)+\nrem(\kappa^{2\gen}-1)} & \text{for } \mgen= \kappa^{2}
\end{cases}
\label{eq_3_11}
\end{equation}
For $\gen \to \infty$, this quantity becomes
\begin{equation}
\langle k^{2}\rangle_{\infty}=
\begin{cases}
\displaystyle \frac{\Krem(\mgen-1)}{\nrem(\mgen-\kappa^{2})}
& \text{for } \mgen> \kappa^{2} \\[8pt]
\infty & \text{for } \mgen\le \kappa^{2}
\end{cases}  .
\label{eq_3_12}
\end{equation}
As will be shown in the next section, the condition for the convergence of
$\langle k^{2}\rangle_{\infty}$ is equivalent to $\gamma>3$, where $\gamma$ is the exponent of the
degree distribution $P(k)\propto k^{-\gamma}$ for $k\gg 1$.

\subsubsection*{Scale-free property}
Let us consider the asymptotic behavior of the degree distribution $P(k)$ of $\mathcal{G}_{\gen}$
for $k\gg 1$ and $\gen\gg 1$. The degree of a node in the $(\gen-1)$-th generation network
$\mathcal{G}_{\gen-1}$ is multiplied by $\kappa$ in $\mathcal{G}_{\gen}$. Therefore, the number of
nodes with degree $k$ in $\mathcal{G}_{\gen-1}$ is identical to the number of nodes with degree
$\kappa k$ in $\mathcal{G}_{\gen}$ if $k$ is larger than the maximum degree of remaining nodes in
$G$, that is,
\begin{equation}
N_{\gen-1}(k)=N_{\gen}(\kappa k) \qquad \text{for }k>\max_{n} [k_{n}^{\text{rem}}]  .
\label{eq_3_13}
\end{equation}
This can be directly confirmed by Eq~(\ref{eq_3_5}). Since it is natural to suppose that the degree
distribution for a large generation converges to a specific functional form $P(k)$, the above
equation is written, in a continuum approximation for $k$, as
\begin{equation}
N_{\gen-1}P(k)d k=N_{\gen}P(\kappa k)d (\kappa k)
\label{eq_3_14}
\end{equation}
for large $\gen$ and $k$. Using Eq~(\ref{eq_3_4}), this relation leads to
\begin{equation}
P(k)=\mgen \kappa P(\kappa k)  ,
\label{eq_3_15}
\end{equation}
and we have the solution of this functional equation as
\begin{equation}
P(k) \propto k^{-\gamma}  ,
\label{eq_3_16}
\end{equation}
where
\begin{equation}
\gamma=1+\frac{\log \mgen}{\log \kappa}  .
\label{eq_3_17}
\end{equation}
This implies that a constructed network possesses the scale-free property and the exponent of the
power-law degree distribution is determined by the degree $\kappa$ of the root node and the number
of edges $\mgen$ in the generator. If the condition (1) for the generator $G$ is violated, the
exponent $\gamma$ diverges and the network $\mathcal{G}_{\gen}$ does not exhibit the scale-free
property. The exponent $\gamma$ always satisfies
\begin{equation}
\gamma \ge 2  ,
\label{eq_3_18}
\end{equation}
because $\mgen\ge 2\kappa$ for any generator. It is easy to check that the exponent $\gamma$ for
the $(u,v)$-flower \cite{Rozenfeld07a}, the SHM model \cite{Song06}, and their derivatives
\cite{Zhang09,Pan10,Hu85,Zhang11b,Gao21} can be reproduced by Eq~(\ref{eq_3_17}).

It should be emphasized that Eq~(\ref{eq_3_16}) does not mean $N_{\gen}(k)\propto k^{-\gamma}$ for
large $t$ and $k$. This is because degrees of nodes in $\mathcal{G}_{\gen}$ can actually take
exponentially discretized values, such as $k$, $\kappa k$, $\kappa^{2}k$, $\dots$, whereas $P(k)$
is defined in the continuum approximation for $k$. The number of nodes of degree $k$ is then given
by $N_{\gen}(k)=N_{\gen}\int_{k}^{\kappa k}P(k')d k'$, which provides the asymptotic behavior of
$N_{\gen}(k)$ as
\begin{equation}
N_{\gen}(k) \propto k^{-\gamma'}  ,
\label{eq_3_19}
\end{equation}
where
\begin{equation}
\gamma'=\gamma-1=\frac{\log \mgen}{\log \kappa}  ,
\label{eq_3_20}
\end{equation}
for a large generation.

Finally, we consider the condition for the convergence (or divergence) of
$\langle k^{2}\rangle_{\infty}$ as shown in Eq~(\ref{eq_3_12}). If $\mgen> \kappa^{2}$,
$\log \mgen /\log \kappa$ is larger than $2$, namely, $\gamma$ given by Eq~(\ref{eq_3_17}) is large
 than $3$. Therefore, $\langle k^{2}\rangle_{\infty}$ becomes finite if $\gamma>3$, while it
diverges for $\gamma \le 3$. This is quite reasonable because of the form of Eq~(\ref{eq_3_16}).

\subsubsection*{Fractal property}

Fractality of a network can be examined by the relation between the number of nodes in the network
and the network diameter, the maximum value of the shortest-path distance, according
to the definition of fractal networks mentioned in the Introduction section. Let
$L_{\gen-1}$ be the diameter of $\mathcal{G}_{\gen-1}$. Then, each of $L_{\gen-1}$ edges between
two nodes separated by the diameter is replaced with the generator $G$ in the network
$\mathcal{G}_{\gen}$ and the shortest-path distance between these two nodes becomes
$\lambda L_{\gen-1}$, where $\lambda$ is the shortest-path distance between the two root nodes in
$G$. If the generation $\gen$ is large enough, the distance $\lambda L_{\gen-1}$ is almost the
longest shortest-path distance in $\mathcal{G}_{\gen}$. More precisely, the diameter of
$\mathcal{G}_{\gen}$ is given by $\lambda L_{\gen-1}+L_{0}$, where $L_{0}$ is a constant. If
$\lambda$ gives the diameter of $G$, we have $L_{0}=0$. Even in the case that the diameter of $G$
is longer than $\lambda$ and $L_{0}$ is finite, however, $\lambda L_{\gen-1}$ is much larger than
$L_{0}$ for $\gen\gg 1$ and we can ignore the term of $L_{0}$. Therefore, the diameter $L_{\gen}$
of $\mathcal{G}_{\gen}$ is expressed as
\begin{equation}
L_{\gen}=\lambda L_{\gen-1}  .
\label{eq_3_21}
\end{equation}
On the other hand, the numbers of nodes $N_{\gen}$ and $N_{\gen-1}$  are related by
Eq~(\ref{eq_3_4}) for $t\gg 1$. Equations (\ref{eq_3_4}) and (\ref{eq_3_21}) implies that if the
network diameter is $\lambda$ times larger, the number of nodes becomes $\mgen$ times larger. This
leads to the relation between $N_{\gen}$ and $L_{\gen}$ as
\begin{equation}
N_{\gen}\propto L_{\gen}^{\df}  ,
\label{eq_3_22}
\end{equation}
where
\begin{equation}
\df=\frac{\log \mgen}{\log \lambda}  .
\label{eq_3_23}
\end{equation}
This result shows that the network $\mathcal{G}_{\gen}$ formed by our model exhibits the fractal
nature with the fractal dimension $\df$ given by Eq~(\ref{eq_3_23}).

The fractal dimensions of the $(u,v)$-flower \cite{Rozenfeld07a} and the SHM model \cite{Song06}
are reproduced by Eq~(\ref{eq_3_23}). If the condition (2) for the generator $G$ is violated,
namely if the two root nodes are directly connected in $G$, the fractal dimension $\df$ diverges
and $\mathcal{G}_{\gen}$ becomes a small-world network.

\subsubsection*{Clustering property}

The clustering property of a network is often characterized by two types of quantities, namely the
average clustering coefficient \cite{Watts98} given by
\begin{equation}
C=\frac{1}{N}\sum_{n=1}^{N} \frac{2\Delta_{n}}{k_{n}(k_{n}-1)}  ,
\label{eq_3_4_1}
\end{equation}
and the global clustering coefficient \cite{Barrat00,Newman01} defined as
\begin{equation}
C^{\triangle}=\frac{\sum_{n}2\Delta_{n}}{\sum_{n} k_{n}(k_{n}-1)}  ,
\label{eq_3_4_2}
\end{equation}
where $N$ is the network size, $k_{n}$ is the degree of node $n$, and $\Delta_{n}$ is the number of
triangles including the node $n$. The quantity $C$ is the average of the local clustering
coefficient $C(n)=2\Delta_{n}/[k_{n}(k_{n}-1)]$, and $C^{\triangle}$ is the ratio of three times
the total number of triangles in the network ($\sum_{n}\Delta_{n}$) to the total number of
connected triplets of nodes [$\sum_{n} k_{n}(k_{n}-1)/2$]. We can calculate analytically these two
clustering coefficients for FSFNs formed by our model.

At first, we derive the average clustering coefficient $C_{\gen}$ of $\mathcal{G}_{\gen}$ in the
generation $\gen$. It should be noted that all triangles in $\mathcal{G}_{\gen}$ are produced in
the edge replacement procedure to form $\mathcal{G}_{\gen}$ from $\mathcal{G}_{\gen-1}$. Newly
added $N_{\gen} -N_{\gen-1}(=\nrem M_{\gen-1})$ nodes in this procedure contribute $\nrem
M_{\gen-1}\crem/N_{\gen}$ to $C_{\gen}$, where $\crem$ is the average clustering coefficients of
the remaining nodes in the generator $G$, i.e.,
\begin{equation}
\crem=\frac{1}{\nrem}\sum_{n\in G}^{\nrem} \frac{2\Delta_{n}}{k_{n}^{\text{rem}}(k_{n}^{\text{rem}}-1)}  .
\label{eq_3_4_3}
\end{equation}
The $N_{\gen-1}$ nodes left in $\mathcal{G}_{\gen}$ are those inherited from
$\mathcal{G}_{\gen-1}$. The degree of a node whose degree in $\mathcal{G}_{\gen-1}$ is $k$ becomes
$\kappa k$ in $\mathcal{G}_{\gen}$, and this node contributes to $k \DR$ triangles in
$\mathcal{G}_{\gen}$, where $\DR$ is the number of triangles in the generator which include one of
the root nodes. Therefore, the average clustering coefficient $C_{\gen}$ of the network
$\mathcal{G}_{\gen}$ is written as
\begin{equation}
C_{\gen}=\frac{1}{N_{\gen}}\left[\nrem M_{\gen-1}\crem+2\DR \sum_{k} \frac{N_{\gen-1}(k)}{\kappa (\kappa k-1)}\right]  .
\label{eq_3_4_4}
\end{equation}
Substituting Eq~(\ref{eq_3_6}) for $N_{\gen-1}(k)$, we can express $C_{\gen}$ by quantities
characterizing the generator. In the limit of $\gen \to \infty$, the average clustering coefficient
becomes
\begin{equation}
C_{\infty}=\frac{\mgen-1}{\mgen}\left[\crem+
\frac{2\DR}{\kappa\nrem}\sum_{n=1}^{\nrem}\sum_{\gen=1}^{\infty}
\frac{1}{\left(\kappa\mgen\right)^{\gen}k_{n}^{\text{rem}}-\mgen^{\gen}} \right]  .
\label{eq_3_4_5}
\end{equation}
This result implies that if there exists even one triangle in the generator, the average clustering
coefficient of $\mathcal{G}_{\gen}$ will be finite, no matter if the size of $\mathcal{G}_{\gen}$
is finite or infinite.

Next, we consider the global clustering coefficient $C^{\triangle}$ defined by
Eq~(\ref{eq_3_4_2}). The denominator $\sum_{n} k_{n}(k_{n}-1)$ is obviously expressed as
$N_{\gen}(\langle k^{2} \rangle_{\gen}-\langle k \rangle_{\gen})$ for $\mathcal{G}_{\gen}$. The
quantity $\sum_{n}\Delta_{n}$ in the numerator is equal to three times the total number of
triangles in the network. Since all triangles in $\mathcal{G}_{\gen}$ are produced in the procedure
of replacing $M_{\gen-1}$ edges in $\mathcal{G}_{\gen-1}$ with the generator $G$, the total number
of triangles in $\mathcal{G}_{\gen}$ is given by $M_{\gen-1}\Dgen$, where $\Dgen$ is the number of
triangles included in $G$. Thus, the global clustering coefficient $C_{\gen}^{\Delta}$ for
$\mathcal{G}_{\gen}$ is given by $C_{\gen}^{\Delta}=6M_{\gen-1}\Dgen/N_{\gen}(\langle k^{2}
\rangle_{\gen}-\langle k \rangle_{\gen})$. Using the relation $M_{\gen-1}/N_{\gen}=\langle
k\rangle_{\gen}/2\mgen$, $C_{\gen}^{\Delta}$ is written as
\begin{equation}
C_{\gen}^{\Delta}=\frac{3\Dgen \langle k\rangle_{\gen}}{\mgen(\langle k^{2} \rangle_{\gen}-\langle k \rangle_{\gen})}  ,
\label{eq_3_4_6}
\end{equation}
where $\langle k \rangle_{\gen}$ and $\langle k^{2} \rangle_{\gen}$ are presented by
Eqs~(\ref{eq_3_7}) and (\ref{eq_3_11}), respectively. In contrast to $C_{\infty}$, even if
$\Dgen\ne 0$, $C_{\infty}^{\Delta}$ is zero if $\mgen\le \kappa^{2}$ (i.e. $\gamma\le 3$), because
$\langle k^{2} \rangle_{\infty}$ diverges.  On the other hand, $C_{\infty}^{\Delta}$ for $\mgen >
\kappa^{2}$ is finite and is, by means of Eqs~(\ref{eq_3_8}) and (\ref{eq_3_12}), expressed as
\begin{equation}
C_{\infty}^{\Delta}=\frac{6(\mgen-\kappa^{2})\Dgen}{\ngen\mgen(\langle k^{2} \rangle_{\text{gen}}-\langle k \rangle_{\text{gen}})}
\qquad \text{for } \mgen>\kappa^{2}  ,
\label{eq_3_4_7}
\end{equation}
where $\langle k \rangle_{\text{gen}}$ and $\langle k^{2} \rangle_{\text{gen}}$ are the average
degree and average squared degree of the generator $G$, respectively.

\subsubsection*{Degree correlation}

The degree correlation between nodes adjacent to each other is described by the joint probability
$P(k,k')$ that one terminal node of a randomly chosen edge has degree $k$ and the other terminal
node has degree $k'$. This probability is expressed by using the number $M(k,k')$ of
$(k,k')$-edges, i.e. edges connecting nodes with degrees $k$ and $k'$, as
\begin{equation}
P(k,k')	=\frac{1+\delta_{k k'}}{2M}M(k,k')  ,
\label{eq_3_5_1}
\end{equation}
where $M$ is the total number of edges in the network. We can derive the joint probability
$P_{\gen}(k,k')$ for $\mathcal{G}_{\gen}$ by counting the number $M_{\gen}(k,k')$ of $(k,k')$-edges
in $\mathcal{G}_{\gen}$. Since all edges in $\mathcal{G}_{\gen}$ are yielded in the replacement of
$M_{\gen-1}$ edges in $\mathcal{G}_{\gen-1}$ with the generator $G$,
$M_{\gen-1}m_{\text{rem}}(k,k')$ edges in $\mathcal{G}_{\gen}$ contribute to $M_{\gen}(k,k')$,
where $m_{\text{rem}}(k,k')$ is the number of $(k,k')$-edges connecting remaining nodes in $G$. In
addition, $M_{\gen}(k,k')$ includes the contribution from edges between nodes inherited from
$\mathcal{G}_{\gen-1}$ and their neighboring nodes. Considering these contributions,
$M_{\gen}(k,k')$ is given by
\begin{eqnarray}
M_{\gen}(k,k')&=&M_{\gen-1}m_{\text{rem}}(k,k') \nonumber \\
&&+\frac{1}{1+\delta_{kk'}}
\left[\frac{k\mu(k')}{\kappa}N_{\gen-1}\left(\frac{k}{\kappa}\right)
     +\frac{k'\mu(k)}{\kappa}N_{\gen-1}\left(\frac{k'}{\kappa}\right)\right]  ,
\label{eq_3_5_2}
\end{eqnarray}
where $\mu(k)$ is the number of nodes with degree $k$ adjacent to one of the root nodes in $G$.
The quantity $N_{\gen-1}(k/\kappa)$ represents the number of nodes inherited from
$\mathcal{G}_{\gen-1}$ whose degrees become $k$ in $\mathcal{G}_{\gen}$. Note that
$N_{\gen-1}(k/\kappa)$ is zero for non-integer $k/\kappa$ and is related to $N_{\gen}(k)$ by
Eq~(\ref{eq_3_5}). Substituting Eq~(\ref{eq_3_5_2}) into Eq~(\ref{eq_3_5_1}), we obtain
\begin{eqnarray}
P_{\gen}(k,k')&=&\frac{1+\delta_{kk'}}{2\mgen}m_{\text{rem}}(k,k') \nonumber \\
&&+\frac{1}{2\mgen^{\gen}}\left[\frac{k\mu(k')}{\kappa}N_{\gen-1}\left(\frac{k}{\kappa}\right)
     +\frac{k'\mu(k)}{\kappa}N_{\gen-1}\left(\frac{k'}{\kappa}\right)\right]  .
\label{eq_3_5_3}
\end{eqnarray}
The first term of Eq (33) gives the contribution from edges between remaining nodes in $G$
accompanying the edge replacement and the second term comes from edges between inherited nodes from
the previous generation and their neighbors. All quantities on the right-hand side of
Eq~(\ref{eq_3_5_3}) are determined from the structure of $G$. It is thus possible to evaluate
analytically the nearest-neighbor degree correlation in the FSFN formed by a given generator. If a
network has no degree correlation, the joint probability must be given by a product of a function
of $k$ and a function of $k'$. Since $P_{\gen}(k,k')$ presented by Eq~(\ref{eq_3_5_3}) cannot be
expressed in the form of such a product, there exists some sort of nearest-neighbor degree
correlation in $\mathcal{G}_{\gen}$. The type of the degree correlation is revealed by various
measures, such as the assortativity \cite{Newman02a}, Spearman's degree rank correlation
coefficient \cite{Litvak13}, and average degree of the nearest-neighbors of nodes with degree $k$
\cite{Pastor-Satorras01}. These measures are calculated by the joint probability $P_{\gen}(k,k')$.

\subsection*{Percolation problem}

A wide choice of generators allows us to prepare a variety of FSFNs. Many indices characterizing
structural features of constructed FSFNs are analytically evaluated as shown in the previous
section. It becomes possible, by employing these FSFNs, to investigate systematically how various
phenomena and dynamics occurring on FSFNs are influenced by the characteristics of networks. The
percolation transition is one of the most fundamental phenomena on complex networks and is deeply
related to the robustness of networks to failure of nodes or edges \cite{Cohen00,Callaway00,Cohen01,Cohen10}
or the spread of disease \cite{Kiss18,Newman02b,Pastor-Satorras15,Wang17}. Although percolation
processes in small-world networks have been extensively studied so far, as reviewed by \cite{Li21},
our understanding of the percolation problem for FSFNs is still limited
\cite{Rozenfeld07b,Rozenfeld10,Hasegawa10,Hasegawa12,Hasegawa13}. For example, the relation between
the percolation threshold and a specific structural feature of FSFN, such as clustering property,
has not yet been systematically studied. In this section, we discuss the percolation transition in
FSFNs formed by our model. We concentrate here on the bond-percolation process corresponding to
edge failure, because the site-percolation problem reflecting node failure is much more
complicated.

We calculate analytically the percolation transition point and some critical exponents for the
bond-percolation problem in which edges are randomly removed from an infinite FSFN
$\mathcal{G}_{\infty}$ with probability $1-p$. For this purpose, it is convenient to define the
\textit{renormalized root nodes} (RRNs) of $\mathcal{G}_{\gen}$ as two nodes corresponding to the
root nodes of the renormalized network (in the sense of the edge renormalization) of
$\mathcal{G}_{\gen}$ by $\mathcal{G}_{\gen-1}$, which takes the same topology as the generator $G$.
Namely, RRNs of $\mathcal{G}_{\gen}$ are the oldest two nodes in $\mathcal{G}_{\gen}$. An example
of RRNs is illustrated by red nodes in Fig~\ref{fig2}B or \ref{fig2}C, which will be referred to
in the section ``Order parameter exponent".

\subsubsection*{Critical point}

For an infinitely large FSFN ($\gen\to \infty$), the shortest-path distance between the RRNs
diverges. Thus, if the RRNs of $\mathcal{G}_{\gen}$ are still connected to each other in a network
$\tilde{\mathcal{G}}_{\gen}(p)$ which is formed by removing edges randomly with probability $1-p$
from $\mathcal{G}_{\gen}$, the network $\tilde{\mathcal{G}}_{\gen}(p)$ is considered to be
percolated. Since the network $\mathcal{G}_{\gen}$ is composed of $\mgen$ pieces of
$\mathcal{G}_{\gen-1}$, the probability $R_{\gen}(p)$ of the RRNs of $\mathcal{G}_{\gen}$ being
connected to each other in $\tilde{\mathcal{G}}_{\gen}(p)$ is related to the probability
$R_{\gen-1}(p)$ of the RRNs of $\mathcal{G}_{\gen-1}$ being connected in
$\tilde{\mathcal{G}}_{\gen-1}(p)$ as
\begin{equation}
R_{\gen}(p)=\pi[R_{\gen-1}(p)]  ,
\label{eq_4_1}
\end{equation}
where $\pi(p)$ is the probability that the two root nodes of the generator $G$ are connected to
each other in a network where edges are randomly removed from $G$ with probability $1-p$. Equation
(\ref{eq_4_1}) has an unstable fixed point at $p=\pc$, i.e. $R_{\gen}(\pc)=R_{\gen-1}(\pc)=\pc$.
The probability $\pc$ gives the percolation critical point, because the percolation probability
$R_{\infty}(\pc)$ is finite. Therefore, the critical point $\pc$ is presented by the non-trivial
and meaningful solution of the equation,
\begin{equation}
\pi(\pc)=\pc  .
\label{eq_4_2}
\end{equation}
We can determine the functional form of $\pi(p)$ for a given generator $G$. The probability
$\pi(p)$ is expressed as
\begin{equation}
\pi(p)=\sum_{m=\lambda}^{\mgen}s_{m}p^{m}(1-p)^{\mgen-m}  ,
\label{eq_4_3}
\end{equation}
where $s_{m}$ is the number of subgraphs of $G$ with $m$ edges in which the root nodes are
connected. The above summation starts from $m=\lambda$, because $s_{m}=0$ for $m<\lambda$. It is
easy to count the number $s_{m}$ by finding numerically such subgraphs from all subgraphs of $G$,
because the total number of subgraphs of $G$ is only $2^{\mgen}$ with $\mgen$ not very large
usually. The function $\pi(p)$ is in general an $\mgen$-th degree polynomial in $p$, but the
polynomial degree can be reduced if there exist edges in $G$ that do not contribute to paths
connecting the root nodes. In such a case, the degree of $\pi(p)$ becomes equal to the number of
edges in $\breve{G}$, where $\breve{G}$ is the core subgraph of $G$ consisting only of all edges
that contribute to paths between the root nodes. The coefficient $s_{m}$ can be computed by
counting the number of subgraphs of $\breve{G}$ instead of $G$. Consequently, FSFNs built from
different generators but with the same core subgraph $\breve{G}$ have the same critical point
$\pc$.

\subsubsection*{Correlation length exponent}

The correlation length (in the sense of the shortest-path distance) of a percolation network
$\tilde{\mathcal{G}}_{\infty}(p)$ near the critical point $\pc$ behaves as
\begin{equation}
\xi=\xi_{0}\left| p-\pc \right|^{-\nu}  ,
\label{eq_4_4}
\end{equation}
where $\nu$ is the critical exponent for the correlation length and $\xi_{0}$ is a constant. When
we renormalize the substrate network $\mathcal{G}_{\infty}$ by the generator $G$, the edge
occupation probability in the renormalized network $\mathcal{G}_{\infty}'$ is given by $\pi(p)$.
Thus, the correlation length of the renormalized percolation network
$\tilde{\mathcal{G}}'_{\infty}[\pi(p)]$ is the same as $\xi$, as expressed by
\begin{equation}
\xi=\xi_{0}'\left| \pi(p)-\pc \right|^{-\nu}  .
\label{eq_4_5}
\end{equation}
The coefficient $\xi_{0}'$ is $\lambda$ times larger than $\xi_{0}$, i.e., $\xi_{0}'=\lambda
\xi_{0}$, because the root nodes in $G$ are separated by $\lambda$. Equations (\ref{eq_4_4}) and
(\ref{eq_4_5}) then lead to
\begin{equation}
\nu=\frac{\log \lambda}{\log \left|\displaystyle \frac{\pi(p)-\pc}{p-\pc}\right|}  .
\label{eq_4_6}
\end{equation}
Therefore, taking the limit $p\to \pc$, the correlation length exponent is given by
\begin{equation}
\nu=\frac{\log \lambda}{\log \pi'(\pc)} ,
\label{eq_4_7}
\end{equation}
where $\pi'(p)$ is the first derivative of $\pi(p)$. As well as the critical probability $\pc$, the
exponent $\nu$ also depends only on the structure of the core subgraph $\breve{G}$ of $G$, because
$\lambda$ is determined by $\breve{G}$.

The correlation volume $N_{\xi}$ is the average number of nodes within the radius $\xi$ from a
node, which is presented by $N_{\xi}=N_{\xi 0}\left| p-\pc \right|^{-\tilde{\nu}}$, where $\tilde{\nu}$
is the correlation volume exponent and $N_{\xi 0}$ is a constant. Similarly to the argument of $\xi$,
the correlation volume of the renormalized percolation network $\tilde{\mathcal{G}}'_{\infty}[\pi(p)]$
is the same as $N_{\xi}$ of $\tilde{\mathcal{G}}_{\infty}(p)$, namely
$N_{\xi}=N_{\xi 0}'\left| \pi(p)-\pc \right|^{-\tilde{\nu}}$. The coefficient $N_{\xi 0}'$ is given by
$N_{\xi 0}'=\mgen N_{\xi 0}$ because of Eq~(\ref{eq_3_4}). These relations give us the exponent
$\tilde{\nu}$ as
\begin{equation}
\tilde{\nu}=\frac{\log \mgen}{\log \pi'(\pc)} .
\label{eq_4_8}
\end{equation}
From Eqs~(\ref{eq_4_7}), (\ref{eq_4_8}), and (\ref{eq_3_23}), we have the simple relation
\begin{equation}
\tilde{\nu}=\df \nu  ,
\label{eq_4_9}
\end{equation}
which is derived also from $N_{\xi}\propto \xi^{\df}$.

\subsubsection*{Order parameter exponent}

Let us consider the critical exponent $\beta$ for the order parameter $P_{\infty}$, i.e., the
probability of a randomly chosen node belonging to the giant connected component in
$\tilde{\mathcal{G}}_{\infty}(p)$. We can calculate $\beta$ by extending the argument for the
$(u,v)$-flower \cite{Rozenfeld07b} to our general model. The order parameter $P_{\infty}$ is the
limiting value of $P_{\gen}(p,N_{\gen})$ for $N_{\gen}\to \infty$, where $P_{\gen}(p,N_{\gen})$ is
the probability that a node in the $\gen$-th generation FSFN $\mathcal{G}_{\gen}$ with $N_{\gen}$
nodes belongs to the largest component in $\tilde{\mathcal{G}}_{\gen}(p)$. According to the
finite-size scaling theory \cite{Stauffer92}, the quantity $P_{\gen}(p,N_{\gen})$ for a large
generation $\gen$ and near the critical point $\pc$ must have the form of
\begin{equation}
P_{\gen}(p,N_{\gen})=N_{\gen}^{-\beta/\tilde{\nu}}F\left[(p-\pc)N_{\gen}^{1/\tilde{\nu}}\right]  ,
\label{eq_4_10}
\end{equation}
where $F(x)$ is a scaling function. At the critical point $p=\pc$, we then have the relation
\begin{equation}
\frac{P_{\gen}(\pc,N_{\gen})}{P_{\gen-1}(\pc,N_{\gen-1})}=\left(\frac{N_{\gen}}{N_{\gen-1}}\right)^{-\beta/\tilde{\nu}}
=\mgen^{-\beta/\tilde{\nu}}  ,
\label{eq_4_11}
\end{equation}
here we used Eq~(\ref{eq_3_4}) for the last equation. Therefore, Eq~(\ref{eq_4_8}) leads to the
expression for the exponent $\beta$ as
\begin{equation}
\beta=-\frac{\log \oc}{\log \pi'(\pc)} ,
\label{eq_4_12}
\end{equation}
where $\oc$ is the ratio of $P_{\gen}(\pc,N_{\gen})$ to $P_{\gen-1}(\pc,N_{\gen-1})$, namely,
\begin{equation}
\oc=\lim_{\gen \to \infty} \frac{P_{\gen}(\pc,N_{\gen})}{P_{\gen-1}(\pc,N_{\gen-1})}  .
\label{eq_4_13}
\end{equation}
Since the probability $\pi(p)$ is presented by Eq~(\ref{eq_4_3}) for a given generator $G$, we can
calculate analytically the exponent $\beta$ from Eq~(\ref{eq_4_12}) if $\oc$ is obtained for $G$.

\begin{figure}[tttt]
\begin{center}
\includegraphics[width=13cm]{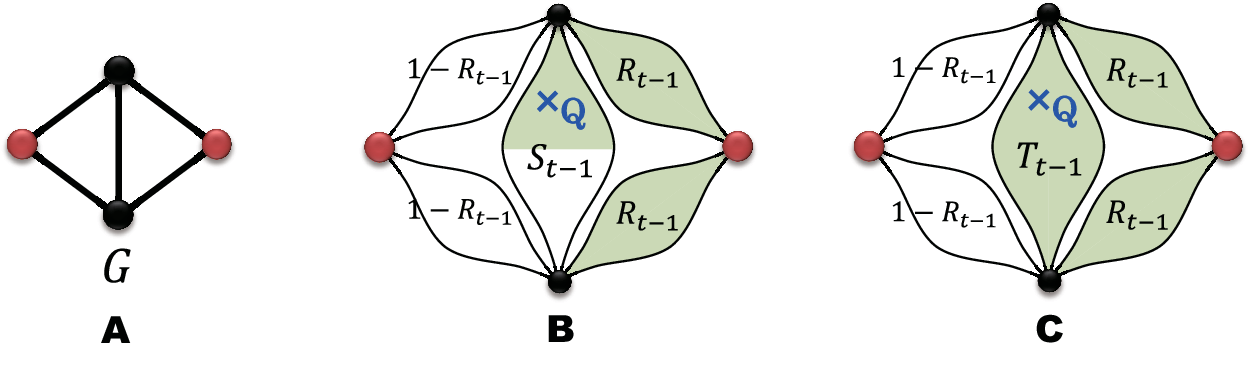}
\caption{{\bf Illustrations showing how a randomly chosen node is connected to one of the two RRNs
in $\boldsymbol{\tilde{\mathcal{G}}_{\gen}}$ with the probability $\boldsymbol{S_{\gen}}$.} (A)
Generator $G$ to construct an FSFN on which the percolation process is considered. The root nodes
are indicated by red circles. Figures (B) and (C) are schematic representations of two percolation
networks on the $\gen$-th generation FSFN $\mathcal{G}_{\gen}$ constructed by $G$. In these
figures, red circles show the two RRNs of $\mathcal{G}_{\gen}$. Green and white eye-shaped parts
represent subgraphs $\tilde{\mathcal{G}}_{\gen-1}(p)$ in which the two RRNs of
$\mathcal{G}_{\gen-1}$ are connected and $\tilde{\mathcal{G}}_{\gen-1}(p)$ with the disconnected
RRNs, respectively. A randomly chosen node Q indicated by a blue cross in (B) is connected to one
of the two RRNs of $\mathcal{G}_{\gen-1}$ (black circle on the top) with the probability
$S_{\gen-1}$ and eventually connected to one of the RRNs of $\mathcal{G}_{\gen}$ (red circle on the
right). In figure (C), however, the node Q is connected to both RRNs of $\mathcal{G}_{\gen-1}$
(black circles on the top and bottom) with the probability $T_{\gen-1}$ and eventually connected to
one of the RRNs of $\mathcal{G}_{\gen}$ (red circle on the right).}
\label{fig2}
\end{center}
\end{figure}

In order to calculate the ratio $\oc$, we introduce two probabilities $S_{\gen}$ and $T_{\gen}$.
The quantity $S_{\gen}$ is the probability that a randomly chosen node Q is connected to one of the
RRNs of $\mathcal{G}_{\gen}$ in the percolation network $\tilde{\mathcal{G}}_{\gen}(p)$, and
$T_{\gen}$ is the probability that the chosen node Q is connected to both RRNs. The probability
$P_{\gen}(p,N_{\gen})$ for a large $\gen$ is then given by
\begin{equation}
P_{\gen}(p,N_{\gen})=S_{\gen}+T_{\gen}  ,
\label{eq_4_14}
\end{equation}
because the probability of the node Q being at a finite distance from either of the RRNs is almost
zero for $\gen\to \infty$. As illustrated in Fig~\ref{fig2}, the probabilities $S_{\gen}$ and
$T_{\gen}$ can be expressed as functions of $S_{\gen-1}$, $T_{\gen-1}$, and the probability
$R_{\gen-1}(p)$ of the two RRNs of $\mathcal{G}_{\gen-1}$ being connected to each other. Since
these functions are linear with respect to $S_{\gen-1}$ and $T_{\gen-1}$, we can express
$(S_{\gen},T_{\gen})$ as
\begin{equation}
\left(
\begin{array}{@{\,}c@{\,}}
S_{\gen} \\
T_{\gen}
\end{array}
\right)
=W
\left(
\begin{array}{@{\,}c@{\,}}
S_{\gen-1} \\
T_{\gen-1}
\end{array}
\right)  ,
\label{eq_4_15}
\end{equation}
where $W$ is a two-by-two matrix whose matrix element $w_{ij}$ with $i,j=1$ or $2$ is a function of
$R_{\gen-1}(p)$. For a large enough generation $\gen$, the largest eigenvalue $\omega$ of the
matrix $W$ gives the ratio $(S_{\gen}+T_{\gen})/(S_{\gen-1}+T_{\gen-1})$ and thus
$P_{\gen}(p,N_{\gen})/P_{\gen-1}(p,N_{\gen-1})$ from Eq~(\ref{eq_4_14}), because the vector
$(S_{\gen},T_{\gen})^{\text{T}}$ becomes proportional to the eigenvector belonging to $\omega$.
Therefore, the ratio $\oc$ defined by Eq~(\ref{eq_4_13}) is simply the largest eigenvalue of
$W$ at $p=\pc$ and for $\gen \to \infty$.
%

Let us determine the functional forms of the matrix elements $w_{ij}[R_{\gen-1}(p)]$ from the
structure of the generator $G$. As seen from the form of Eq~(\ref{eq_4_15}), the element $w_{11}$
is, for example, the conditional probability that a randomly chosen node Q is connected to one of
the two RRNs of $\mathcal{G}_{\gen}$ under the condition that the node Q is connected to one of the
RRNs of the subgraph $\mathcal{G}_{\gen-1}$ including Q. Various connection patterns contribute to
this conditional probability $w_{11}$. A situation illustrated in Fig~\ref{fig2}B also contributes
to $w_{11}$. Since the probability of this situation occurring is
$R_{\gen-1}^{2}[1-R_{\gen-1}]^{2}S_{\gen-1}$, this connection pattern is incorporated into $w_{11}$
as a term of $R_{\gen-1}^{2}[1-R_{\gen-1}]^{2}$. By the same token, $w_{12}$ contains a
contribution from a situation shown in Fig~\ref{fig2}C. As demonstrated by these examples, the
matrix element $w_{ij}$ is expressed as a polynomial consisting of terms of
$R_{\gen-1}^{m}[1-R_{\gen-1}]^{\mgen-m-1}$ with $0\le m\le \mgen-1$. Since $R_{\gen}(\pc)$ becomes
equal to $\pc$ for $\gen \to \infty$, the matrix element $w_{ij}$ in the thermodynamic limit at
$p=\pc$ is presented by
\begin{equation}
w_{ij}=\frac{1}{\mgen(1+\delta_{1j})}\sum_{m=0}^{\mgen-1}c_{ij}(m)\pc^{m}(1-\pc)^{\mgen-m-1}  ,
\label{eq_4_16}
\end{equation}
where $c_{ij}(m)$ is the number of connection patterns, with $m$ connected subgraphs
$\tilde{\mathcal{G}}_{\gen-1}(\pc)$, that a randomly chosen node Q is connected to RRNs of
$\mathcal{G}_{\gen}$. The prefactor $1/\mgen$ for $j=2$ is the probability that the node Q is
included in a specific subgraph $\mathcal{G}_{\gen-1}$, and $1/2\mgen$ for $j=1$ is the probability
that the node Q is in a subgraph $\mathcal{G}_{\gen-1}$ and one of the RRNs of
$\mathcal{G}_{\gen-1}$ is chosen as the connection point of Q. In order to calculate the
coefficient $c_{ij}(m)$ from the structure of $G$, we consider subgraphs $G'(e_{0})$ formed by
removing single edges $e_{0}$ from $G$. The edge $e_{0}$ corresponds to the subgraph
$\tilde{\mathcal{G}}_{\gen-1}(\pc)$ including the node Q. The coefficient $c_{ij}(m)$ is the number
of subgraphs of $G'(e_{0})$ for any $e_{0}$ with $m$ edges, in which $j(=1 \text{ or } 2)$ terminal
nodes of $e_{0}$ are connected to $i(=1 \text{ or } 2)$ root nodes of $G$ under the condition that
only for $j=2$ the terminal nodes of $e_{0}$ are considered to be connected directly to each other
even though the edge $e_{0}$ is absent in $G'(e_{0})$. Because of the small size of $G$, the
numbers of these subgraphs can be counted numerically, as in the case of the evaluation of $s_{m}$
in Eq~(\ref{eq_4_3}). We can eventually obtain the order parameter exponent $\beta$ from
Eq~(\ref{eq_4_12}) by calculating the largest eigenvalue $\oc$ of the matrix $W$ whose elements are
given by Eq~(\ref{eq_4_16}).

\subsubsection*{Examples}

\begin{figure}[tttt]
\begin{center}
\includegraphics[width=11cm]{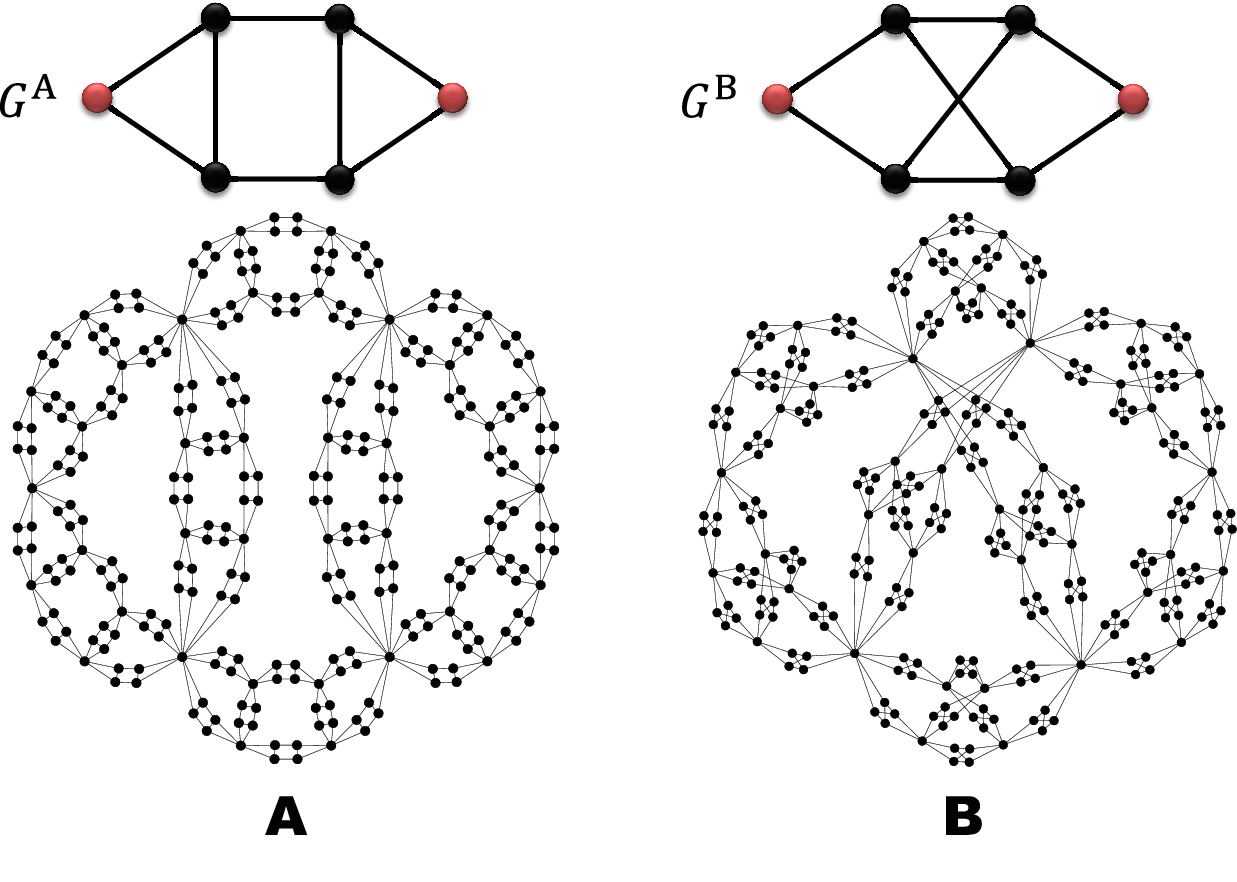}
\caption{{\bf Two generators $\boldsymbol{G^{\text{A}}}$ and $\boldsymbol{G^{\text{B}}}$ (upper
panel) and the 3rd generation FSFNs $\boldsymbol{\mathcal{G}_{3}^{\text{A}}}$ and
$\boldsymbol{\mathcal{G}_{3}^{\text{B}}}$ formed by these generators (lower panel).} The root nodes
in $G^{\text{A}}$ and $G^{\text{B}}$ are represented by red circles. Since both generators have the
same numbers of nodes ($\ngen=6$) and edges ($\mgen=8$), the numbers of nodes and edges in
$\mathcal{G}_{3}^{\text{A}}$ and $\mathcal{G}_{3}^{\text{B}}$ are also the same, namely $N_{3}=294$
and $M_{3}=512$.
}
\label{fig3}
\end{center}
\end{figure}

As examples of the above general argument, let us demonstrate the structural features of FSFNs
formed by two kinds of generators, and compare the critical properties of percolation on these
networks. We employ two generators $G^{\text{A}}$ and $G^{\text{B}}$ shown in Figs~\ref{fig3}A and
\ref{fig3}B, respectively. These generators have the same number of nodes ($\ngen=6$), number of
edges ($\mgen=8$), degree of the root node ($\kappa=2$), shortest-path distance between the root
nodes ($\lambda=3$), and degrees of remaining nodes ($k_{n}^{\text{rem}}=3$ for any $n$). The
similarity found in the generators $G^{\text{A}}$ and $G^{\text{B}}$ leads to similar structural
features of the FSFNs constructed by them. In fact, the FSFNs $\mathcal{G}_{\infty}^{\text{A}}$ and
$\mathcal{G}_{\infty}^{\text{B}}$ formed by $G^{\text{A}}$ and $G^{\text{B}}$, respectively, in the
infinite generation possess the same average degree $\langle k\rangle_{\infty}=7/2$ calculated by
Eq~(\ref{eq_3_8}), same second moment $\langle k^{2}\rangle_{\infty}=63/4$ by Eq~(\ref{eq_3_12}),
same scale-free property $\gamma=4$ by Eq~(\ref{eq_3_17}), and same fractal dimension $\df=\log
8/\log 3=1.893$ by Eq~(\ref{eq_3_23}). In addition, the joint probabilities $P_{\gen}(k,k')$ given
by Eq~(\ref{eq_3_5_3}), which describe the nearest-neighbor degree correlation, are also the same
for $\mathcal{G}_{\gen}^{\text{A}}$ and $\mathcal{G}_{\gen}^{\text{B}}$ for any $\gen$, because the
nearest-neighbor degree correlations in $G^{\text{A}}$ and $G^{\text{B}}$ are the same. As a
result, the Spearman's degree rank correlation coefficient becomes $\varrho=-21/64$ for both FSFNs
in the infinite generation.
%

Clustering properties of these two FSFNs are, however, different from each other. Since there is no
triangle in $G^{\text{B}}$, the average clustering coefficient $C_{\gen}$ and the global clustering
coefficient $C^{\triangle}_{\gen}$ are zero for any generation FSFN formed by $G^{\text{B}}$. On
the contrary, the generator $G^{\text{A}}$ has two triangles and the local clustering coefficients
of all nodes in $G^{\text{A}}$ are finite. Therefore, both of the two kinds of clustering
coefficients $C_{\gen}$ and $C^{\triangle}_{\gen}$ take finite values. The average clustering
coefficient of $\mathcal{G}_{\infty}^{\text{A}}$ is $C_{\infty}=0.31486$ which is obtained from
Eq~(\ref{eq_3_4_5}) with $\DR=1$, $C_{\text{rem}}=4/3$, and $k_{n}^{\text{rem}}=3$ for any $n$. We
can also calculate the global clustering coefficient of $\mathcal{G}_{\infty}^{\text{A}}$ as
$C^{\triangle}_{\infty}=3/14$ from Eq~(\ref{eq_3_4_7}) with $\triangle_{\text{gen}}=2$.

We compare properties of percolation on $\mathcal{G}_{\infty}^{\text{A}}$ and
$\mathcal{G}_{\infty}^{\text{B}}$. Counting numerically the number of subgraphs of $G^{\text{A}}$,
the coefficients $s_{m}$ for various $m$ in Eq~(\ref{eq_4_3}) are calculated as $s_{3}=2$,
$s_{4}=14$, $s_{5}=34$, $s_{6}=25$, $s_{7}=8$, and $s_{8}=1$ for $G^{\text{A}}$. Similarly, we have
$s_{3}=4$, $s_{4}=20$, $s_{5}=40$, $s_{6}=26$, $s_{7}=8$, and $s_{8}=1$ for $G^{\text{B}}$. Thus,
the functions $\pi(p)$ for $G^{\text{A}}$ and $G^{\text{B}}$ have different forms of 8th order
polynomials. The coefficients $c_{ij}(m)$ in Eq~(\ref{eq_4_16}) for the matrix elements $w_{ij}$
are also obtained by counting numerically the numbers of subgraphs of the generators satisfying
required conditions. The largest eigenvalues of the matrix $W$ are then computed as $\oc=0.9649$
for $G^{\text{A}}$ and $0.9344$ for $G^{\text{B}}$. These quantities characterizing the generators,
such as $\pi(p)$, $\oc$, $\mgen$, and $\lambda$, give the percolation critical point $\pc$ and
critical exponents for bond percolation on $\mathcal{G}_{\infty}^{\text{A}}$ and
$\mathcal{G}_{\infty}^{\text{B}}$ as shown in Table \ref{tab1}. The validity of these values, as
well as structural measures, has been confirmed by numerical calculations. Although most of the
structural features of $\mathcal{G}_{\infty}^{\text{A}}$ and $\mathcal{G}_{\infty}^{\text{B}}$ are
the same, except for the clustering property, $\mathcal{G}_{\infty}^{\text{B}}$ is more robust than
$\mathcal{G}_{\infty}^{\text{A}}$ against edge elimination and the percolation transitions on these
FSFNs belong to different universality classes with different critical exponents. As can be seen
from these examples, our generalized model enables us to examine systematically the relationship
between certain structural properties of FSFNs, such as the clustering property, and phenomena or
dynamics on them.
%
\begin{table}[ttt]
\centering
\caption{
{\bf \label{tab1}Values of the percolation critical point $\boldsymbol{\pc}$ and critical exponents
$\boldsymbol{\nu}$, $\boldsymbol{\tilde{\nu}}$, and $\boldsymbol{\beta}$ for FSFNs formed by the
generators $\boldsymbol{G^{\text{A}}}$ and $\boldsymbol{G^{\text{B}}}$.}}
\begin{tabular}{l|l|l|l|l}
\hline
\rowcolor[rgb]{0.9,0.9,0.9}
\textbf{Generator} & $\boldsymbol{p}_{\textbf{c}}$ & $\boldsymbol{\nu}$ & $\boldsymbol{\tilde{\nu}}$ & $\boldsymbol{\beta}$  \\ \hline
$G^{\text{A}}$     & $0.6961$                      & $1.8293$           & $3.4626$                   & $0.0595$ \\ \hline
$G^{\text{B}}$     & $0.6288$                      & $1.7772$           & $3.3638$                   & $0.1098$ \\ \hline
\end{tabular}
\label{table1}
\end{table}

\subsection*{Asymmetric generator}
\begin{figure}[tttt]
\begin{center}
\includegraphics[width=10cm]{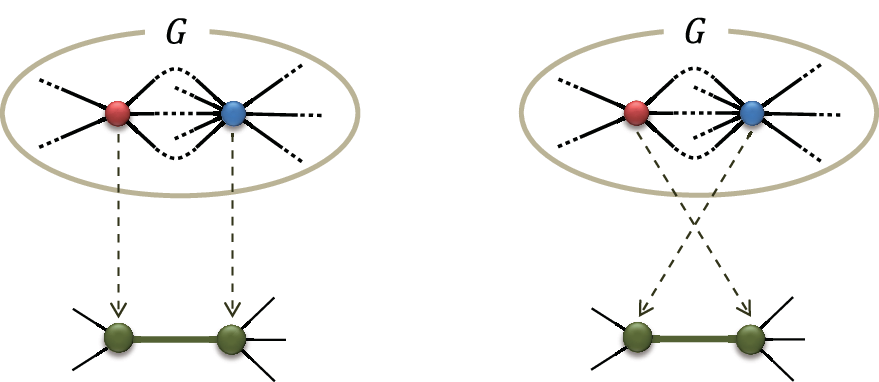}
\caption{{\bf Two ways to replace an edge with an asymmetric generator $\boldsymbol{G}$.} Which
replacement way is chosen is random with the probability $1/2$. The red and blue circles indicate
the non-equivalent root nodes in the asymmetric generator.
}
\label{fig4}
\end{center}
\end{figure}

Up to here, a generator must satisfy the conditions (1), (2), and (3) as mentioned in the Model
section. The condition (3) guarantees that the resulting FSFN has a deterministic structure. We
consider in this section the case that the condition (3) is violated, namely, the generator is
{\it asymmetric}. An asymmetric generator $G$ is a network whose subgraph obtained by removing one
root node and its edges from $G$ has a different topology from a network obtained by removing
another root node and its edges from $G$. Since the two root nodes of an asymmetric generator are
not equivalent, there are two ways to replace an edge with the generator as illustrated by
Fig~\ref{fig4}. Here, we assume that the way of edge replacement is randomly chosen with the
probability $1/2$. This stochasticity makes a final network $\mathcal{G}_{\gen}$ non-deterministic,
but $\mathcal{G}_{\gen}$ is still fractal and has the scale-free property in a statistical sense,
as will become clear in the discussion below. We show how the various results obtained for
symmetric generators are modified by the asymmetry of generators.
%

The numbers of nodes $N_{\gen}$ and edges $M_{\gen}$ are given by Eqs~(\ref{eq_3_2}) and
(\ref{eq_3_3}), respectively, as in the case of symmetric generators, because these quantities are
not influenced by the symmetry of generators. The number of nodes of degree $k$ is, however,
different from Eq~(\ref{eq_3_6}). The probability that $k_{1}$ edges from a node of degree $k$ are
replaced with the generator $G$ in one way and the remaining $(k-k_{1})$ edges are replaced
with $G$ in another way is given by $\Big( \begin{array}{@{}c@{}} k \\[-5pt] k_{1} \end{array} \Big)/2^{k}$.
Considering the probability that $k_{1}$ edges from a node of degree $k'$ in the $(\gen-1)$-th
generation network $\mathcal{G}_{\gen-1}$ are replaced with $G$ in one specific way, the number of
nodes $N_{\gen}(k)$ of degree $k$ in $\mathcal{G}_{\gen}$ is presented by
\begin{equation}
N_{\gen}(k)=\sum_{k'=1}\frac{N_{\gen-1}(k')}{2^{k'}}\sum_{k_{1}=0}^{k'}
\left(
\begin{array}{@{}c@{}}
k' \\
k_{1}
\end{array}
\right)
\delta_{k,\kappa_{1}k_{1}+\kappa_{2}(k'-k_{1})}
+M_{\gen-1}\sum_{n=1}^{\nrem}\delta_{k,k_{n}^{\text{rem}}}  ,
\label{eq_5_1}
\end{equation}
where $\kappa_{1}$ and $\kappa_{2}$ are the degrees of the two root nodes of $G$. If
$\kappa_{1}=\kappa_{2}$, the above relation becomes equivalent to Eq~(\ref{eq_3_5}). We can obtain
$N_{\gen}(k)$ by solving the recurrence Eq~(\ref{eq_5_1}) with the initial condition
$N_{1}(k)=\delta_{k \kappa_{1}}+\delta_{k \kappa_{2}}+\sum_{n=1}^{\nrem}\delta_{k,k_{n}^{\text{rem}}}$.
The moments $\langle k\rangle_{\gen}$ and $\langle k^{2}\rangle_{\gen}$ can be calculated from this
$N_{\gen}(k)$. The average degree $\langle k\rangle_{\gen}$ is given by Eq~(\ref{eq_3_7}) [or (\ref{eq_3_8})
for $\gen\to \infty$], because $N_{\gen}$ and $M_{\gen}$ are unchanged from those for symmetric generators.
Computing $\langle k^{2}\rangle_{\gen}=\sum_{k}k^{2}N_{\gen}(k)/N_{\gen}$, the second moment in the infinite
generation limit is expressed as
\begin{equation}
\langle k^{2}\rangle_{\infty}=
\begin{cases}
\displaystyle \frac{(\mgen-1)(2\check{\kappa}^{2}+\Krem)}{\nrem(\mgen-\bar{\kappa}^{2})}
& \text{for } \mgen> \bar{\kappa}^{2}\\[8pt]
\infty & \text{for } \mgen\le \bar{\kappa}^{2}
\end{cases}  ,
\label{eq_5_2}
\end{equation}
where
\begin{eqnarray}
\bar{\kappa} &=& \frac{\kappa_{1}+\kappa_{2}}{2}  ,\label{eq_5_3}\\
\check{\kappa} &=& \frac{|\kappa_{1}-\kappa_{2}|}{2}  .\label{eq_5_4}
\end{eqnarray}
It is obvious that Eq~(\ref{eq_5_2}) coincides with Eq~(\ref{eq_3_12}) if $\kappa_{1}=\kappa_{2}$.

In order to examine the scale-free property of $\mathcal{G}_{\gen}$, let us consider the asymptotic
behavior of $N_{\gen}(k)$ for large values of $\gen$ and $k$. Neglecting stochastic fluctuations in
edge replacements taken into account in Eq~(\ref{eq_5_1}), one can find the asymptotic form of
$N_{\gen}(k)$. In an average sense, $k/2$ edges from a node of degree $k$ in $\mathcal{G}_{\gen-1}$
are multiplied by $\kappa_{1}$ and the remaining $k/2$ edges are multiplied by $\kappa_{2}$ in
$\mathcal{G}_{\gen}$. Therefore, we have
\begin{equation}
N_{\gen-1}(k)=N_{\gen}(\bar{\kappa} k) \qquad \text{for } k>\max_{n} [k_{n}^{\text{rem}}]  .
\label{eq_5_5}
\end{equation}
This expression is the same as Eq~(\ref{eq_3_13}) with $\bar{\kappa}$ instead of $\kappa$. Thus,
according to the argument deriving Eq~(\ref{eq_3_17}) from Eq~(\ref{eq_3_13}) in the section
``Scale-free property", the degree distribution $P(k)$ obeys asymptotically $P(k)\propto
k^{-\gamma}$, where the exponent $\gamma$ is given by
\begin{equation}
\gamma=1+\frac{\log \mgen}{\log \bar{\kappa}}  .
\label{eq_5_6}
\end{equation}
This implies that a network $\mathcal{G}_{\gen}$ formed by an asymmetric generator also exhibits
the scale-free property. It is interesting to note that $\mathcal{G}_{\gen}$ is scale-free as long
as $\bar{\kappa}$ is greater than $1$, even if the degree of one of the root nodes is unity,
regardless of the condition (1) for generators. Here, we should emphasize that the relation between
$N_{\gen}(k)$ and $P(k)$ depends on the values of $\kappa_{1}$ and $\kappa_{2}$. If
$\kappa_{1}=\kappa_{2}$ holds, degrees in  $\mathcal{G}_{\gen}$ take exponentially discretized
values, such as $k$, $\bar{\kappa} k$, $\bar{\kappa}^{2}k$, $\dots$, as in the case of symmetric
generators, and $N_{\gen}(k)$ is proportional to $kP(k)$, i.e. $N_{\gen}(k)\propto k^{-\gamma'}$
with $\gamma'=\log \mgen/\log \bar{\kappa}$. On the other hand, degrees in $\mathcal{G}_{\gen}$
take nearly uniform values if $\kappa_{1}\ne \kappa_{2}$. In this case, the exponent $\gamma'$
becomes equal to $\gamma$, because $N_{\gen}(k)\propto P(k)$. Therefore, the exponent $\gamma'$
describing the asymptotic behavior of $N_{\gen}(k)$ is presented by
\begin{equation}
\gamma'=
\begin{cases}
\displaystyle \frac{\log \mgen}{\log \bar{\kappa}}     & \text{for } \kappa_{1}=\kappa_{2} \\[12pt]
\displaystyle 1+\frac{\log \mgen}{\log \bar{\kappa}}   & \text{for } \kappa_{1}\ne\kappa_{2}
\end{cases} .
\label{eq_5_7}
\end{equation}

The fractal property of a network $\mathcal{G}_{\gen}$ is not influenced by the symmetry of the
generator. This is because the relation between the diameter $L_{\gen-1}$ of the $(\gen-1)$-th
generation network $\mathcal{G}_{\gen-1}$ and $L_{\gen}$ for $\mathcal{G}_{\gen}$ is still
expressed by Eq~(\ref{eq_3_21}) even for the asymmetric generator, and the number of nodes
$N_{\gen}$ also remains as given by Eq~(\ref{eq_3_3}). Therefore, $\mathcal{G}_{\gen}$ formed by an
asymmetric generator keeps the fractal property with the fractal dimension given by
Eq~(\ref{eq_3_23}).

In order to obtain the expression of the average clustering coefficient of the FSFN formed by an
asymmetric generator, we need to distinguish the numbers of triangles $\DRa$ and $\DRb$ including
the root nodes $\text{R}_{1}$ and $\text{R}_{2}$, respectively. The probability that $k_{1}\DRa$
triangles arise from the replacement of $k_{1}$ edges from a node of degree $k$ with the generator
$G$ and $(k-k_{1})\DRb$ triangles arise from the replacement of the remaining $(k-k_{1})$ edges
is $\Big( \begin{array}{@{}c@{}} k \\[-5pt] k_{1} \end{array} \Big)/2^{k}$. Considering this
probability, the average clustering coefficient is presented by
\begin{eqnarray}
C_{\gen}&=&\frac{1}{N_{\gen}}\Bigg[\nrem M_{\gen-1}\crem \nonumber \\
&& + \left. \sum_{k}\sum_{k_{1}=0}^{k} \frac{N_{\gen-1}(k)}{2^{k-1}}
\left( \begin{array}{@{}c@{}} k \\ k_{1} \end{array} \right)
\frac{\DRa k_{1}+\DRb (k-k_{1})}{h(k,k_{1})[h(k,k_{1})-1]} \right]  ,
\label{eq_5_8}
\end{eqnarray}
where $h(k,k_{1})=\kappa_{1}k_{1}+\kappa_{2}(k-k_{1})$ and $N_{\gen-1}(k)$ is given by
Eq~(\ref{eq_5_1}). This corresponds to Eq~(\ref{eq_3_4_4}) for symmetric generators. Since the
total numbers of triangles and connected triplets in the network $\mathcal{G}_{\gen}$ are not
affected by the symmetry of the generator, the global clustering coefficient $C_{\gen}^{\Delta}$ is
presented by Eq~(\ref{eq_3_4_6}). In the thermodynamic limit ($\gen\to\infty$),
$C_{\infty}^{\Delta}=0$ for $\mgen \le \bar{\kappa}^{2}$ and $C_{\infty}^{\Delta}$ is given by
Eq~(\ref{eq_3_4_7}) with $\bar{\kappa}$ instead of $\kappa$ if $\mgen > \bar{\kappa}^{2}$.

In the symmetric generator case, only nodes with degree $k/\kappa$ in $\mathcal{G}_{\gen-1}$ can
become nodes with degree $k$ in $\mathcal{G}_{\gen}$. The joint probability $P_{\gen}(k,k')$ is
then provided by Eq~(\ref{eq_3_5_3}). For FSFNs by asymmetric generators, however, all nodes with
degree $k''$ in $\mathcal{G}_{\gen-1}$ can be nodes with degree $k$ in $\mathcal{G}_{\gen}$ if
$k''$ satisfies $h(k'',k_{1})=k$ for arbitrary integer $k_{1} \in [0,k'']$. Taking into account the
probability of choosing $k_{1}$ edges from $k''$ edges, we can write the joint probability
$P_{\gen}(k,k')$ as
\begin{eqnarray}
P_{\gen}(k,k')&=&\frac{1+\delta_{kk'}}{2\mgen}m_{\text{rem}}(k,k') \nonumber \\
&+& \frac{1}{2\mgen^{\gen}}\sum_{k''}\sum_{k_{1}=0}^{k''}\frac{N_{\gen-1}(k'')}{2^{k''}}
\left( \begin{array}{@{}c@{}} k'' \\ k_{1} \end{array} \right) \left[J_{k''k_{1}}(k,k')+J_{k''k_{1}}(k',k)\right] ,
\label{eq_5_9}
\end{eqnarray}
where $J_{k''k_{1}}(k,k')=[k_{1}\mu_{1}(k)+(k''-k_{1})\mu_{2}(k)]\delta_{h(k'',k_{1}),k'}$ and
$\mu_{1(2)}(k)$ is the number of nodes with degree $k$ adjacent to the root node $\text{R}_{1(2)}$
in $G$.

The percolation process on our FSFN depends only on how the two root nodes are connected to each
other in the generator $G$, irrespective of the symmetry of $G$. Thus, the critical point and
critical exponents of the bond-percolation transition on FSFNs formed by asymmetric generators are
also obtained by the same argument as that for symmetric generators. Although the renormalization
of $\mathcal{G}_{\gen}$ by $\mathcal{G}_{\gen-1}$ is not uniquely determined in the asymmetric
generator case because of the structural fluctuation in these networks, $\mathcal{G}_{\gen}$ can be
renormalized into the network with the same topology as the generator $G$ by using any of the
various realizations of $\mathcal{G}_{\gen-1}$. The RRNs of $\mathcal{G}_{\gen}$ or its subgraph
$\mathcal{G}_{\gen-1}$ by which $\mathcal{G}_{\gen}$ is renormalized are also defined as the nodes
corresponding to the root nodes of the renormalized network. Eventually, the critical point $\pc$
is given by the solution of Eq~(\ref{eq_4_2}), and the critical exponents $\nu$, $\tilde{\nu}$, and
$\beta$ are provided by Eqs~(\ref{eq_4_7}), (\ref{eq_4_8}), and (\ref{eq_4_12}), respectively. The
only difference from the case of the symmetric generator is that the numbers of subgraphs $s_{m}$
and $c_{ij}(m)$ are calculated for the asymmetric generator $G$.

\begin{figure}[tttt]
\begin{center}
\includegraphics[width=0.9\textwidth]{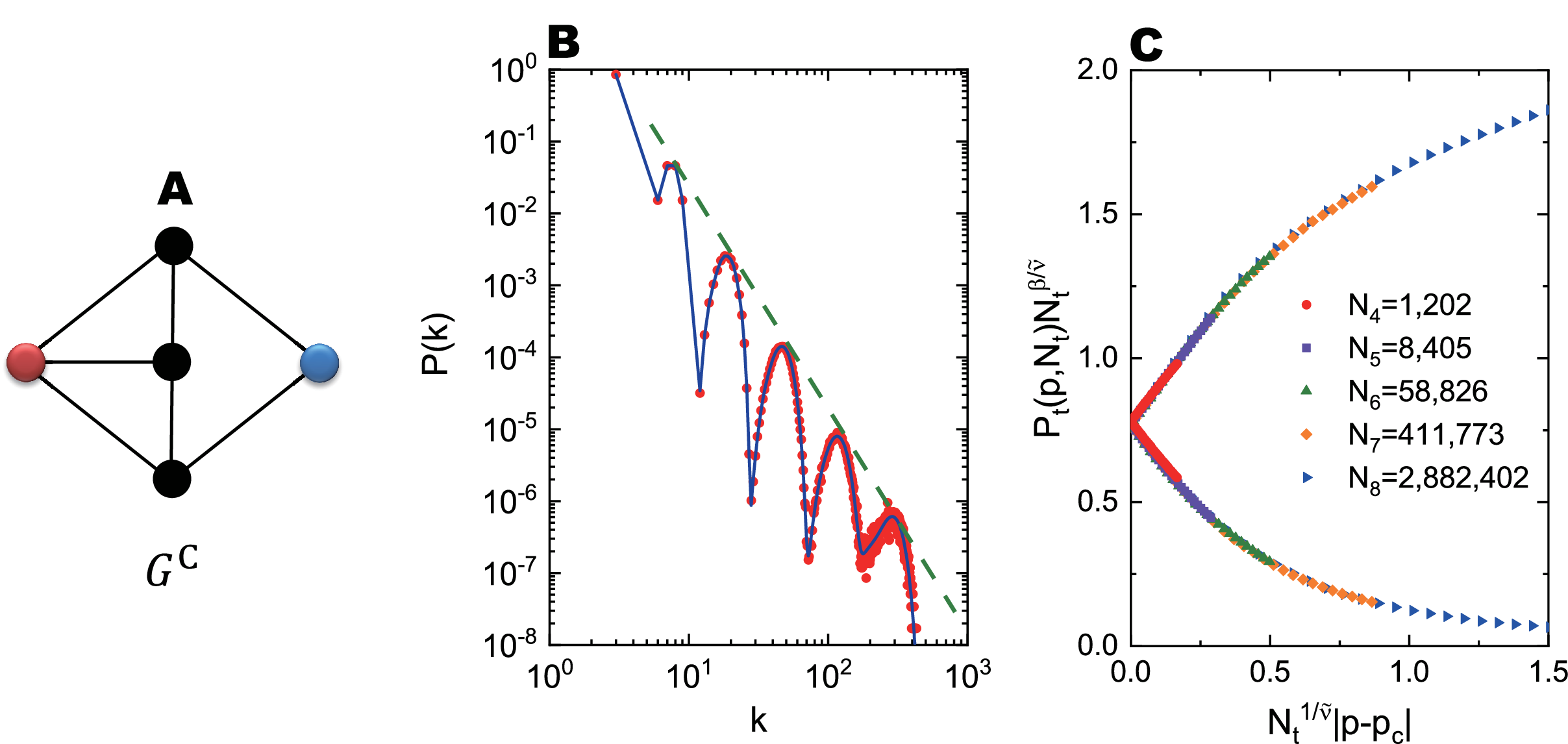}
\caption{{\bf Degree distribution and the percolation property of the FSFN formed by an asymmetric
generator.} (A) Example of an asymmetric generator $G^{\text{C}}$. The red and blue circles
represent the root nodes of $G^{\text{C}}$. (B) Degree distribution $P(k)$ for the $6$th generation
FSFN $\mathcal{G}_{6}^{\text{C}}$ formed by $G^{\text{C}}$. The network $\mathcal{G}_{6}^{\text{C}}$
contains $N_{6}=58,826$ nodes and $M_{6}=117,649$ edges. The solid blue line shows the analytically
calculated $P(k)$ and red symbols represent numerical result averaged over $1,000$ realizations of
FSFNs. The dashed line gives the slope of $P(k)\propto k^{-\gamma}$ with $\gamma=3.1237$ obtained by
Eq~(\ref{eq_5_6}). (C) Scaling plot of the order parameter $P_{\gen}(p,N_{\gen})$ for
$\mathcal{G}_{\gen}^{\text{C}}$ in various generations by using the analytical values of $\pc$,
$\tilde{\nu}$, and $\beta$. The network sizes $N_{\gen}$ of FSFNs are shown in the figure.
}
\label{fig5}
\end{center}
\end{figure}

As an example, let us consider the FSFN $\mathcal{G}_{\gen}^{\text{C}}$ formed by the asymmetric
generator $G^{\text{C}}$ illustrated by Fig~\ref{fig5}A. Basic measures characterizing
$G^{\text{C}}$ are $\mgen=7$, $\nrem=3$, $\kappa_{1}=3$, and $\kappa_{2}=2$. Since $\kappa_{1}\ne
\kappa_{2}$, the degree distribution $P(k)$ of $\mathcal{G}_{\gen}^{\text{C}}$ is simply obtained
by $N_{\gen}(k)/N_{\gen}$ and is analytically calculated by using Eqs~(\ref{eq_3_3}) and
(\ref{eq_5_1}). The solid line in Fig~\ref{fig5}B represents the theoretical result of $P(k)$ for
the $6$th generation FSFN $\mathcal{G}_{6}^{\text{C}}$. Symbols showing the numerical result agree
well with this theoretical curve. The peaks equally spaced on the logarithmic $k$-axis reflect the
nested appearance of the binomial distribution in $N_{\gen}(k)$ given by Eq~(\ref{eq_5_1}).
Equation (\ref{eq_5_6}) gives the power-law exponent describing $P(k)$ as $\gamma=3.1237$. The
envelope of $P(k)$ surely exhibits this power-law behavior as shown in Fig~\ref{fig5}B. The average
degree and average squared degree of $\mathcal{G}_{\infty}^{\text{C}}$ are $\langle
k\rangle_{\infty}=4$ and $\langle k^{2}\rangle_{\infty}=220/3$, respectively. The fractal dimension
of $\mathcal{G}_{\infty}^{\text{C}}$ is $\df=\log 7/\log 2 \approx 2.8074$. Since the generator
contains triangles and $\mgen>\bar{\kappa}^{2}$, the network $\mathcal{G}_{\infty}^{\text{C}}$ is
clustered in both senses of $C_{\infty}$ and $C_{\infty}^{\Delta}$, as is clear from the fact that
$C_{\infty}=0.3952$ and $C_{\infty}^{\Delta}=9/182\approx 0.04945$. The degree correlation of
$\mathcal{G}_{\gen}^{\text{C}}$ can be evaluated by the joint probability $P_{\gen}(k,k')$ which is
computed by Eq~(\ref{eq_5_9}). We can calculate the assortativity $r_{\gen}$ \cite{Newman02a} and
Spearman's degree rank correlation coefficient $\varrho_{\gen}$ \cite{Litvak13} from
$P_{\gen}(k,k')$. Although $r_{\gen}$ is zero for $\gen \to \infty$ because $\gamma<4$ and then
$\langle k^{3}\rangle_{\infty}=\infty$, the Spearman's correlation coefficient $\varrho_{\gen}$ is
finite for any $\gen$. In the thermodynamic limit, this coefficient is calculated as
$\varrho_{\infty}=-0.5221$, which indicates that $\mathcal{G}_{\infty}^{\text{C}}$ exhibits
disassortative degree correlations between neighboring nodes. Percolation properties of
$\mathcal{G}_{\infty}^{\text{C}}$ can be also clarified by adapting the argument in the section
``Percolation problem" for symmetric generators. The critical point and critical exponents for the
bond-percolation process on $\mathcal{G}_{\infty}^{\text{C}}$ are calculated as $\pc=0.4473$,
$\nu=1.2561$, $\tilde{\nu}=3.5262$, and $\beta=0.2761$ from Eqs~(\ref{eq_4_2}), (\ref{eq_4_7}),
(\ref{eq_4_8}), and (\ref{eq_4_12}), respectively. The validity of these results is demonstrated in
Fig~\ref{fig5}C that is a scaling plot depicting $P_{\gen}(p,N_{\gen})N_{\gen}^{\beta/\tilde{\nu}}$
as a function of $N_{\gen}^{1/\tilde{\nu}}|p-\pc |$ for $\mathcal{G}_{\gen}^{\text{C}}$ in various
generations $\gen$. According to Eq~(\ref{eq_4_10}), the fact that these plots for different
$N_{\gen}$ fall on the same curve, as shown in Fig~\ref{fig5}C, implies that the calculated $\pc$,
$\tilde{\nu}$, and $\beta$ are correct values. It should be emphasized that data points for each
size in figure \ref{fig5}C are obtained from a single realization of
$\mathcal{G}_{\gen}^{\text{C}}$ and $P_{\gen}(p,N_{\gen})$ is almost independent of samples for
large $\gen$. This is because the global connectivity of $\mathcal{G}_{\gen}$ is irrelevant to the
way of edge replacements.
%

\section*{Conclusion and discussion}
In this work, we have proposed a general structural model of fractal scale-free networks (FSFNs)
and calculated analytically various measures characterizing structures of constructed networks. As
an example of analyses of phenomena occurring on our FSFNs, the percolation problem on infinite
FSFNs has been studied. Using the present model, one can provide a wide variety of deterministic
and non-deterministic FSFNs which include those formed by existing models and examine
systematically the influence of a specific structural property on a phenomenon on FSFNs.

To construct an FSFN, we first prepare a small graph called a generator $G$ in which two particular
nodes are specified as root nodes. The degrees of the root nodes must be no less than $2$, and the
shortest-path distance between the two root nodes have to be $2$ or longer. An FSFN in the
$\gen$-th generation, $\mathcal{G}_{\gen}$, is formed by replacing every edge in the previous
generation network $\mathcal{G}_{\gen-1}$ with the generator $G$ iteratively so that the terminal
nodes of the edge coincide with the root nodes of $G$. If the generator $G$ is symmetric with
respect to the root nodes, the constructed network $\mathcal{G}_{\gen}$ is deterministic, and vice
versa. The most distinct advantage of this model is that a generator $G$ can be chosen arbitrarily
and this enables us to control the scale-free property, fractality, and other structural properties
of FSFNs. Using topological information of $G$, we have analytically presented various indices and
quantities that describe the structure of the FSFN. The obtained analytical expressions ensure that
these quantities can be changed independently by varying the structure of the generator $G$.

We have also studied the bond-percolation problem on infinite FSFNs built by our model and computed
analytically the critical point $\pc$ and various critical exponents. Furthermore, the effect of
the clustering property on the percolation transition has been examined by comparing the critical
points of FSFNs whose structural properties are the same as each other except for the clustering
coefficient. As demonstrated by this example, the present model makes it possible to elucidate how
a specific structural property influences a phenomenon occurring on FSFNs by varying systematically
the structures of FSFNs.

The present model builds an FSFN by replacing every edge with a single specific generator. This
model can be extended to a model in which two or more generators are employed. Networks formed by
such an extended model will keep the fractal and scale-free properties. In a model with two
generators, for example, an edge is replaced with a generator $G_{1}$ with the probability $p$ or
with another generator $G_{2}$ with the probability $1-p$. It is easy to show that the exponent
$\gamma$ is presented by Eq~(\ref{eq_3_17}) with $\langle \mgen\rangle$ and $\langle \kappa\rangle$
instead of $\mgen$ and $\kappa$, respectively, where $\langle \mgen\rangle$ is the mean number of
edges and $\langle \kappa\rangle$ is the mean degree of the root nodes of the multiple generators.
In the two-generator model, these mean quantities are simply given by
$\langle \mgen\rangle=p m_{\text{gen 1}}+(1-p) m_{\text{gen 2}}$ and
$\langle \kappa\rangle=p\kappa_{1}+(1-p)\kappa_{2}$, where $m_{\text{gen 1(2)}}$ is the number of
edges in $G_{1(2)}$ and $\kappa_{1(2)}$ is the degree of the root node of $G_{1(2)}$. If the
generators are asymmetric, $\kappa_{1(2)}$ is the average degree of the two root nodes in
$G_{1(2)}$. The fractal dimension $\df$ is also written as Eq~(\ref{eq_3_23}) with $\langle \mgen\rangle$
and $\langle \lambda\rangle=p \lambda_{1}+(1-p)\lambda_{2}$, where $\lambda_{1(2)}$ is the shortest-path
distance between the root nodes in $G_{1(2)}$. Other measures characterizing the constructed
network $\mathcal{G}_{\gen}$ are computed in similar ways to the calculations for a single
asymmetric generator. Since these measures are continuous functions of the probability that a
generator is adopted for an edge replacement, we can control more freely and finely the structural
properties of $\mathcal{G}_{\gen}$ by adjusting the adoption probability. The idea of constructing
a network by means of mixed or probabilistic edge replacements with two kinds of small graphs has
already been considered in the SHM model and the extension of the $(u,v)$-flower model, though the
obtained network is not fractal \cite{Song06,Diggans20,Ma20}. The above extended model can be
regarded as a generalization of this idea. The extension to multiple generators does not just
provide a highly controllable mathematical model. The multi-generator model could be relevant to
the formation mechanism of real-world FSFNs. As seen in the growth process of the World Wide Web or
trading networks, many real networks grow by replacing their constituent elements with small motifs
or hierarchical combinations of them. The multi-generator model suggests that networks become
fractal and scale-free if the replacing procedure satisfies some conditions. Therefore, the present
model and its extensions open up avenues for a systematic understanding of phenomena occurring on
FSFNs and for the elucidation of formation mechanisms of real-world FSFNs.

\section*{Acknowledgments}
The authors thank T.~Kitahara for fruitful discussions. This work was supported by a Grant-in-Aid
for Scientific Research (Grant No.~19K03646) from the Japan Society for the Promotion of Science
and by Moonshot Research and Development Program (No.~JPMJMS2023) from the Japan Science and
Technology Agency.

%
%
%

\end{document}